% ...

\documentclass[preprint]{aastex}

\newcommand{\tnm}{\tablenotemark}
\newcommand{\tnt}{\tablenotetext}
\newcommand{\ch}{\colhead}
\newcommand{\mc}{\multicolumn}
\newcommand{\Av}{$\rm{A}_V$}
\newcommand{\Msun}{M$_\odot$}

\newcommand{\p}{$\pm$~}
\newcommand{\logg}{$\log (g)$}
\newcommand{\Teff}{$T_{\rm{eff}}$}
\newcommand{\kt}{KPNO Tau 3}
\newcommand{\Spitzer}{\textit{Spitzer}}
\newcommand{\Herschel}{\textit{Herschel}}

\newcommand{\tco}{$^{13}$CO}
\newcommand{\ceo}{C$^{18}$O}
\newcommand{\chisq}{$\chi^2$}
\newcommand{\amax}{$a_{max}$}
\newcommand{\amin}{$a_{min}$}

\usepackage{graphicx}
\usepackage{epsfig}
\usepackage{epstopdf}
\usepackage{natbib}
\usepackage{color}

\shorttitle{Brown Dwarf Disk in Taurus}
\shortauthors{Broekhoven-Fiene, Matthews, Duch{\^e}ne et al.}

\begin{document}

\title{The disk around the brown dwarf KPNO Tau 3}

\author{Hannah Broekhoven-Fiene\altaffilmark{1}, Brenda Matthews\altaffilmark{1,2}, Gaspard Duch{\^e}ne\altaffilmark{3,4}, 
James Di Francesco\altaffilmark{1,2}, 
Aleks Scholz\altaffilmark{5},
Antonio Chrysostomou\altaffilmark{6}, 
and Ray Jayawardhana\altaffilmark{7}
}

\altaffiltext{1}{Department of Physics and Astronomy, University of Victoria, Victoria, BC, V8W 3P6, Canada}
\altaffiltext{2}{Herzberg Institute of Astrophysics, National Research Council of Canada, Victoria, BC, V9E 2E7, Canada}
\altaffiltext{3}{Department of Astronomy, University of California at Berkeley, Hearst Field Annex, B-20, Berkeley CA 94720-3411, USA}
\altaffiltext{4}{UJF-Grenoble 1/CNRS-INSU, Institut de Plan{\'e}tologie et d'Astrophysique (IPAG), UMR 5274, F-38041 Grenoble, France}
\altaffiltext{5}{School of Physics \& Astronomy, University of St Andrews, North Haugh, St Andrews, KY16 9SS, UK}
\altaffiltext{6}{Joint Astronomy Centre, 660 North A{\'o}hoku Place, University Park, Hilo, HI 96720, USA}
\altaffiltext{7}{Department of Astronomy and Astrophysics, University of Toronto, 50 St. George Street, Toronto, ON M5S 3H4, Canada}

\begin{abstract}
We present submillimeter observations of the young brown dwarfs KPNO Tau 1, KPNO Tau 3, and KPNO Tau 6 at 450 \micron\ and 850 \micron\ taken with the Submillimeter Common-User Bolometer Array on the James Clerke Maxwell Telescope. \kt\ and KPNO Tau 6 have been previously identified as Class II objects hosting accretion disks, whereas KPNO Tau 1 has been identified as a Class III object and shows no evidence of circumsubstellar material. Our  3 $\sigma$ detection of cold dust around \kt\ implies a total disk mass of (4.0 $\pm$ 1.1) $\times~10^{-4}$ \Msun\ (assuming a gas to dust ratio of 100:1). We place tight constraints on any disks around KPNO Tau 1 or KPNO Tau 6 of $<2.1 \times10^{-4}$ \Msun~and $<2.7 \times10^{-4}$ \Msun, respectively. Modeling the spectral energy distribution of \kt\ and its disk suggests the disk properties (geometry, dust mass, and grain size distribution) are consistent with observations of other brown dwarf disks and low-mass T-Tauri stars. In particular, the disk-to-host mass ratio for \kt\ is congruent with the scenario that at least some brown dwarfs form via the same mechanism as low-mass stars.
\end{abstract}


\section{Introduction}
\label{sec:intro_kt3}

An outstanding question regarding the formation of brown dwarfs (and very low mass stars), is whether the processes are scaled down versions of star-formation processes (e.g., turbulent fragmentation of molecular clouds and cores; \citealt{PadoanNordlund2004}) or whether the accretion of the material onto the brown dwarf is halted (e.g., by ejection of the stellar embryo from its environment whether a massive circumstellar disk or a forming cluster, \citealt{ReipurthClarke2001}, by processes such as photoionization by nearby OB stars or tidal shears within clusters). (See \citealt{Luhman2012} for a recent review on the formation and evolution of brown dwarfs and very low mass stars and the observational constraints.) A number of these different formation scenarios may occur, but the detection of circumsubstellar accretion disks around brown dwarfs, with properties similar to T-Tauri stars, suggests that at least some brown dwarfs go through a T-Tauri like stage and therefore form via similar processes as stars.

Our brown dwarf targets are located in the nearby Taurus star-forming region \citep[distance of 140 pc $\pm$ 10 pc;][]{1994Kenyon}, where the initial mass function has been shown to extend well below the substellar mass limit \citep{Bricenoetal2002}. The detection of accretion onto brown dwarfs \citep{JayawardhanaMohantyBasri2003,Barrado2004} and circumsubstellar disks \citep{Jayawardhanaetal2003} in Taurus suggested that brown dwarfs may undergo a T-Tauri-like phase. Subsequent studies of these disks revealed more similarities between disks around young brown dwarfs and disks around young stars, such as their disk-to-host mass ratios \citep{ScholzJayaWood2006}, and the disk scale heights and flaring angles \citep{Harveyetal2012}. Recent observations have also shown evidence of grain growth in brown dwarf disks to \micron\ and millimeter sizes (e.g., \citealt{Apaietal2004,Apaietal2005,Bouyetal2008,Riccietal2012}), a process shown to take place within T Tauri disks. One difficulty in characterizing brown dwarf disk masses is that many observations at long wavelengths result in non-detections. \cite{Mohantyetal2013} used previous submillimeter and millimeter observations, detections and upper limits, along with new SCUBA-2 observations, to investigate protoplanetary disk masses across stellar and substellar regimes. Observations of brown dwarf disks with millimeter interferometers, such as the Submillimeter Array (SMA), CARMA, and ALMA (e.g., \citealt{Riccietal2012, Riccietal2013,Andrewsetal2013}), have helped overcome some of the sensitivity issues involved in detecting these faint disks.

Brown dwarf disk properties can be used to constrain formation scenarios of brown dwarfs and of objects within their disks. For example, a truncated disk can reveal whether or not a brown dwarf likely formed as a result of the ejection of a stellar embryo from its environment \citep{Umbreitetal2011}. Brown dwarf disk properties also reveal the potential for planet formation within their disks \citep{PayneLodato2007}. Furthermore, spectral slopes at submillimeter wavelengths and longer, which are sensitive to the grain size distribution, can be used to probe grain growth to \micron\ and millimeter sizes, the earliest stage of planet formation in the core-accretion model \citep{Pollacketal1996}.

We present here submillimeter observations of three brown dwarfs, KPNO Tau 1, KPNO Tau 3 and KPNO Tau 6 (of spectral types M8.5, M6 and M8.5, respectively: \citealt{Bricenoetal2002}), in the Taurus star-forming region, to estimate their respective disk masses. Observations at submillimeter and millimeter wavelengths probe the optically thin dust emission and thus are much more effective at deriving the total dust mass. These observations were originally part of a larger survey to investigate brown dwarf disks using the Submillimetre Common-User Bolometer Array (SCUBA; \citealt{Holland1999}) on the James Clerk Maxwell Telescope (JCMT). Only these three targets were observed, however, before SCUBA was decommissioned in 2005. KPNO Tau 3 and KPNO Tau 6 have Class II spectral energy distributions (SEDs), indicating the presence of a circumstellar disk, whereas KPNO Tau 1 has a Class III SED \citep{Hartmannetal2005,Luhmanetal2010}, showing no evidence of circumstellar material. Accretion signatures have been detected from KPNO Tau 3 and KPNO Tau 6 \citep{Barrado2004,JayawardhanaMohantyBasri2005}. These disks have also been observed with the \textit{Herschel Space Observatory}: KPNO Tau 6 was detected by \cite{Harveyetal2012} and KPNO Tau 3 was detected as part of another \Herschel\ program (Bulger et al., submitted). KPNO Tau 3 has also recently been observed with the SMA \citep{Andrewsetal2013}. We observed these disks with SCUBA to measure their masses and compare them to known relations for young low-mass stars and brown dwarfs.

In Section~\ref{sec:obs_kt3}, we report the observations that were taken at the JCMT. The results are described in Section~\ref{sec:results_kt3} where we describe photospheric models (Section~\ref{sec:phot_kt3}) of the brown dwarfs, the disk mass measurements from the dust emission (Section~\ref{sec:mass_kt3}), and \tco\ and \ceo\ observations of KPNO Tau 3 (Section~\ref{sec:massMolecules_kt3}). We present a disk model for KPNO Tau 3 in Section~\ref{sec:model}. Finally, we discuss and summarize our findings in Section~\ref{sec:conclusions}.


\section{Observations and Data Reduction}  
\label{sec:obs_kt3}

\subsection{Photometry with SCUBA}

\begin{deluxetable}{ccccccccc}
\tabletypesize{\scriptsize}
\tablewidth{0pc}
\tablecaption{Observing Log
\label{tbl:log}}
\tablehead{
\ch{Target} 	& \ch{R.A.}		& \ch{Decl.} 	& \ch{Date}		& \ch{Integration}	& \mc{2}{c}{No. of Noisy Bolometers\tnm{a}}	& \mc{2}{c}{Flux Calibration Factor} \\ 
\ch{} 		& \ch{}			& \ch{}			& \ch{Observed}	& \ch{Time}			& \ch{450 \micron}	& \ch{850 \micron}		& \ch{450 \micron}	& \ch{850 \micron} \\
\ch{} 		& \ch{(J2000)}	& \ch{(J2000)}	& \ch{}			& \ch{(s)}			& \ch{}				& \ch{}					& \ch{(Jy/Volt)}		& \ch{(Jy/Volt)}} 
\startdata
KPNO Tau 1	& 04:15:14.71	& +28:00:09.6	& 2004 Sept 13 	& 4705				& 36					& 8						& 379 $\pm$ 11 		& 243 $\pm$ 2\\
KPNO Tau 3	& 04:26:29.39	& +26:24:13.8	& 2004 Sept 13	& 2326				& 36					& 8						& 379 $\pm$ 11 		& 243 $\pm$ 2\\
			&				& 				& 2004 Sept 17	& 2325				& 33					& 9						& 379 $\pm$ 11 		& 243 $\pm$ 2\\
KPNO Tau 6	& 04:30:07.24	& +26:08:20.8	& 2004 Oct 18	& 2299				& 24					& 14						& 480 $\pm$ 60 		& 221 $\pm$ 6
\enddata
\tnt{a}{There are a total of 91 bolometers at 450 \micron\ and 37 bolometers at 850 \micron.}
\tablecomments{The quality of the weather severely degraded during the night of the October observations. Therefore, the fiducial FCFs for the epoch of the observations are used.}
\end{deluxetable}

Photometry observations were taken with SCUBA on the JCMT in 2004 September and October at 850 \micron\  and 450 \micron\ and are summarized in Table~\ref{tbl:log}. The data were reduced using the SCUBA User Reduction Facility  \citep[SURF:][]{1998ASPC..145..216J, 1998SPIE.3357..548J}.

The atmospheric extinction was determined using measurements from the Caltech Submillimeter Observatory (CSO) taumeter at 225 GHz at 10 minute intervals as skydips before and after the observations were not always available. The extinction correction was done using existing relations to extrapolate the CSO measurements to the extinction at the SCUBA bands using the well-established relations from the JCMT \citep{Archibald02}. (Using the CSO tau values for the correction also resulted in better signal-to-noise values than using skydip extinction measurements, where available.) It should be noted that the noise is higher in the 450 \micron\ \kt\  data from the first night. This difference is likely because those data were taken at the end of the night and therefore through more atmosphere. The 450 \micron\ data were affected more strongly as they are more sensitive to atmospheric opacity.

The central bolometer was used for photometry observations of the targets and the median of the remaining bolometers was used to characterize and remove the sky signal. Bolometers that proved to be noisy at any point during the night were not used (see Table~\ref{tbl:log}).

The flux calibration factors (FCFs) used to calibrate the absolute flux scale are given in Table~\ref{tbl:log}. As is typical, we adopt a flux uncertainty of $\sim$20\%. Observations of Uranus were used to measure the FCF for the September observations. The same FCF was used for both nights (four nights apart) as Uranus was only observed on the second night. This extrapolation is reasonable since the predicted flux of Uranus changed very little between the two nights and the measured mean flux of  KPNO Tau 3 varied little between the two nights after FCF correction. (Furthermore, the measured FCF values are consistent with the fiducial FCFs for the epoch of the observations.) Although Uranus was observed multiple times during the night of the October observations, there was a sharp increase in the atmospheric extinction mid-shift, making it difficult to measure FCFs. Thus, the fiducial FCFs for the epoch of the observations are used instead.

\begin{deluxetable}{cccc}
\tablewidth{0pc}
\tablecaption{Flux Measurements and Disk Masses
\label{tbl:kt_disks}}
\tablehead{
\ch{Target}	& \ch{$F_{450}$}	& \ch{$F_{850}$}	& \ch{$M_{\rm{disk}}$} 		\\ 
\ch{}		& \ch{(mJy)}		& \ch{(mJy)}		& \ch{($10^{-4}$$M_\odot$)}}
\startdata
KPNO Tau 1	& $<$25			& $<$3.7			& $<$2.1			\\
KPNO Tau 3	& 48 $\pm$ 18	& 5.9 $\pm$ 2.2	& 4.0 $\pm$ 1.1	\\
KPNO Tau 6	& $<$22			& $<$8.1			& $<$2.7
\enddata
\tablecomments{2 $\sigma$ upper limits are quoted for KPNO Tau 1 and KPNO Tau 6.}
\end{deluxetable}

Our criterion for a detection is the measurement of a non-zero mean that is at least 3$\sigma$, where $\sigma$ is the statistical error after clipping.\footnote{Individual data points more than 3$\sigma$ from the raw data mean are clipped.} With this criterion, only KPNO Tau 3 had a significant flux detection at 450 \micron\ or 850 \micron. The observations of KPNO Tau 3 from each of the two nights independently have a 2$\sigma$ detection at 850 \micron. Concatenating the 850 \micron\ data yields a 3$\sigma$ detection. Conversely, including the 450 \micron\ data from the first night degrades the signal. (The 450 \micron\ data from the first night were highly affected by larger atmospheric extinction, see discussion above). Therefore, we only use the 450 \micron\ data from 2004 17 September for subsequent analysis. Table~\ref{tbl:kt_disks} summarizes the submillimeter data for KPNO Tau 1, KPNO Tau 3, and KPNO Tau 6. We calculate the spectral index of the SED between 450 \micron\ and 850 \micron\ to be $\alpha = 3.3 \pm 1.1$, where $F_\nu \propto \nu^{\alpha}$.


\subsection{Spectroscopy with Receiver A3}

\begin{deluxetable}{ccccccc}
\tablewidth{0pc}
\tablecaption{\label{tbl:spectra} RxA3 Data}
\tablehead{
\ch{Molecular Line}	& \ch{Frequency}	& \ch{Noise}	& \ch{Peak}			& \ch{Line Width}  \\ 
\ch{} 				& \ch{(GHz)}  	& \ch{(K)} 	& \ch{(K)}			& \ch{(km/s)}}
\startdata
\tco\ $J=2-1$ 			& 220.399  		& 0.011 		& 0.50 \p 0.02		& 0.35 \p 0.02 \\
\ceo\ $J=2-1$ 			& 219.560  		& 0.011 		& $<$0.022		& \nodata 
\enddata
\end{deluxetable}

Follow-up spectroscopy was done using Director's Discretionary Time on the JCMT on 2011 January 25 at 220 GHz. The \tco\ $J=2-1$ and \ceo\ $J=2-1$ lines were observed simultaneously. The science observation details are summarized in Table~\ref{tbl:spectra}. Data were reduced using the Sub-Millimetre User Reduction Facility \citep[SMURF: ][]{SMURF} and the VO-enabled Spectral Analysis Tool \citep[SPLAT: ][]{SPLAT}.

Baselines were fit to regions of the spectra that did not contain the spectral line of interest or the noisy ends of the spectra. The noise in each spectrum is listed with the target molecular lines in Table~\ref{tbl:spectra}.

\begin{figure}
\includegraphics[trim=3.7cm 16cm 3.2cm 3.5cm, clip=true, width=4.65 in]{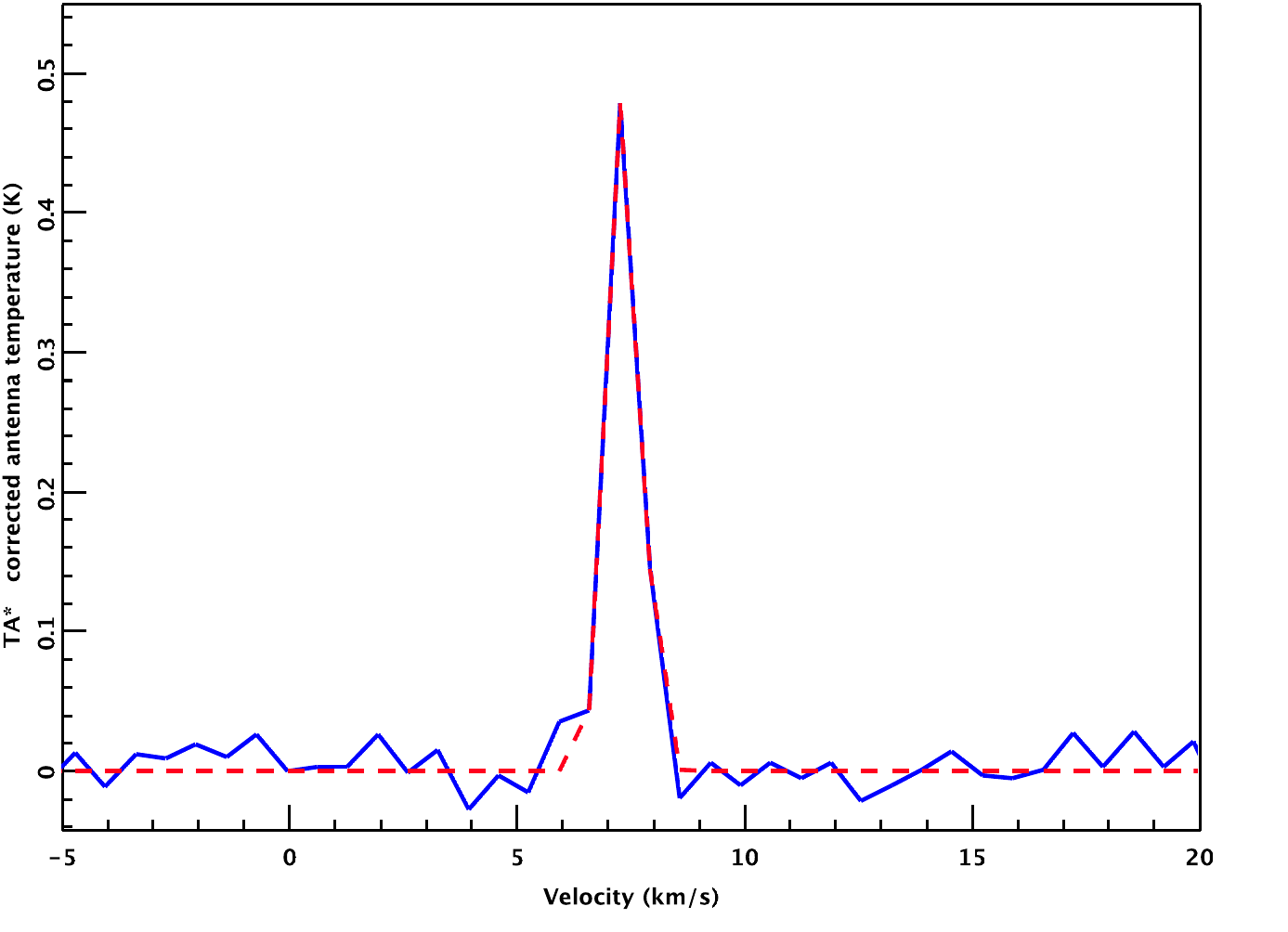}\centering\\
\includegraphics[trim=3.7cm 23.5cm 3.2cm 3.13cm, clip=true, width=4.485 in]{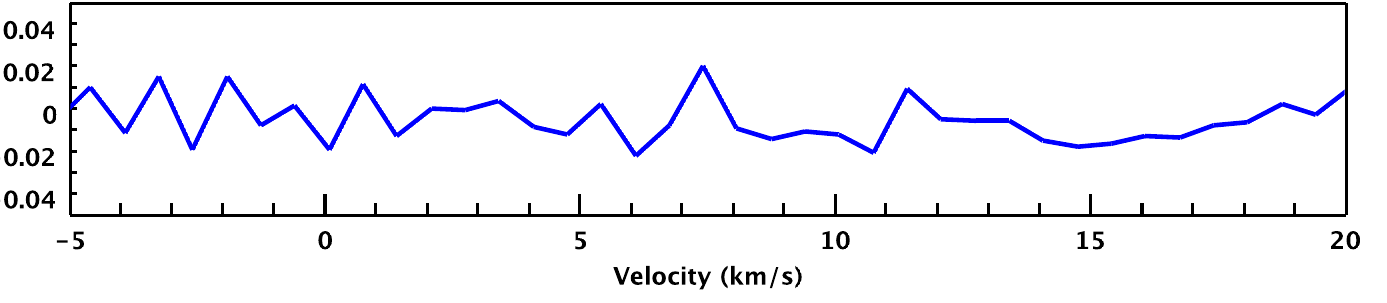}
\caption{\label{fig:spectra}
Spectral regions containing \tco\ $J=2-1$ (top) and \ceo\ $J=2-1$ (bottom) lines are shown. The rest velocity for each panel is set to the frequency of the respective molecular line (therefore the top and bottom panels correspond to different frequency ranges.) The top panel shows the Gaussian profile (red dashed line) that is fit to the observed spectrum (blue solid line). No spectral line is detected for \ceo\ at 219.56 GHz.\scriptsize}
\end{figure}
 
The \tco\ and \ceo\ spectra are shown in Figure~\ref{fig:spectra}. A \tco\ spectral line is observed and fit with a Gaussian profile to measure the peak brightness, 0.50 $\pm$ 0.01 K, and line width, $\Delta\nu$ = 0.61 $\pm$ 0.01 MHz (0.35 $\pm$ 0.02 km/s compared to a channel width of 0.66 km/s). There is no line detected in the \ceo\ spectrum and so the 2 $\sigma$ upper limit of 0.022 K on the peak brightness and the \tco\ line width are adopted for the analysis.


\section{Results}
\label{sec:results_kt3}

\subsection{Modeling of Photospheric Emission}
\label{sec:phot_kt3}

\begin{figure}
\includegraphics[scale=0.45, clip=True, trim=0.4cm 0.2cm 2cm 1.2cm]{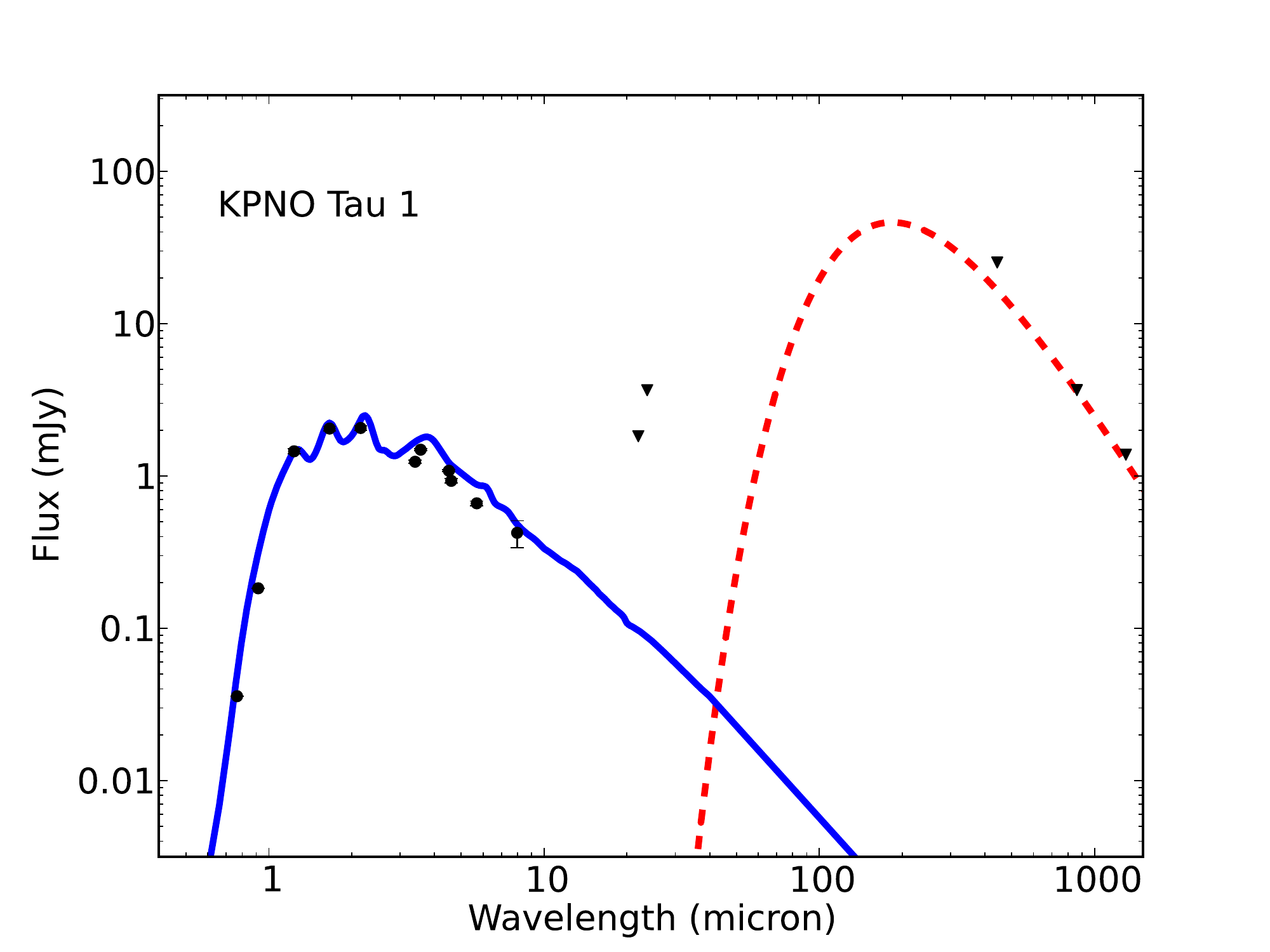}\centering\\
\includegraphics[scale=0.45, clip=True, trim=0.4cm 0.2cm 2cm 1.2cm]{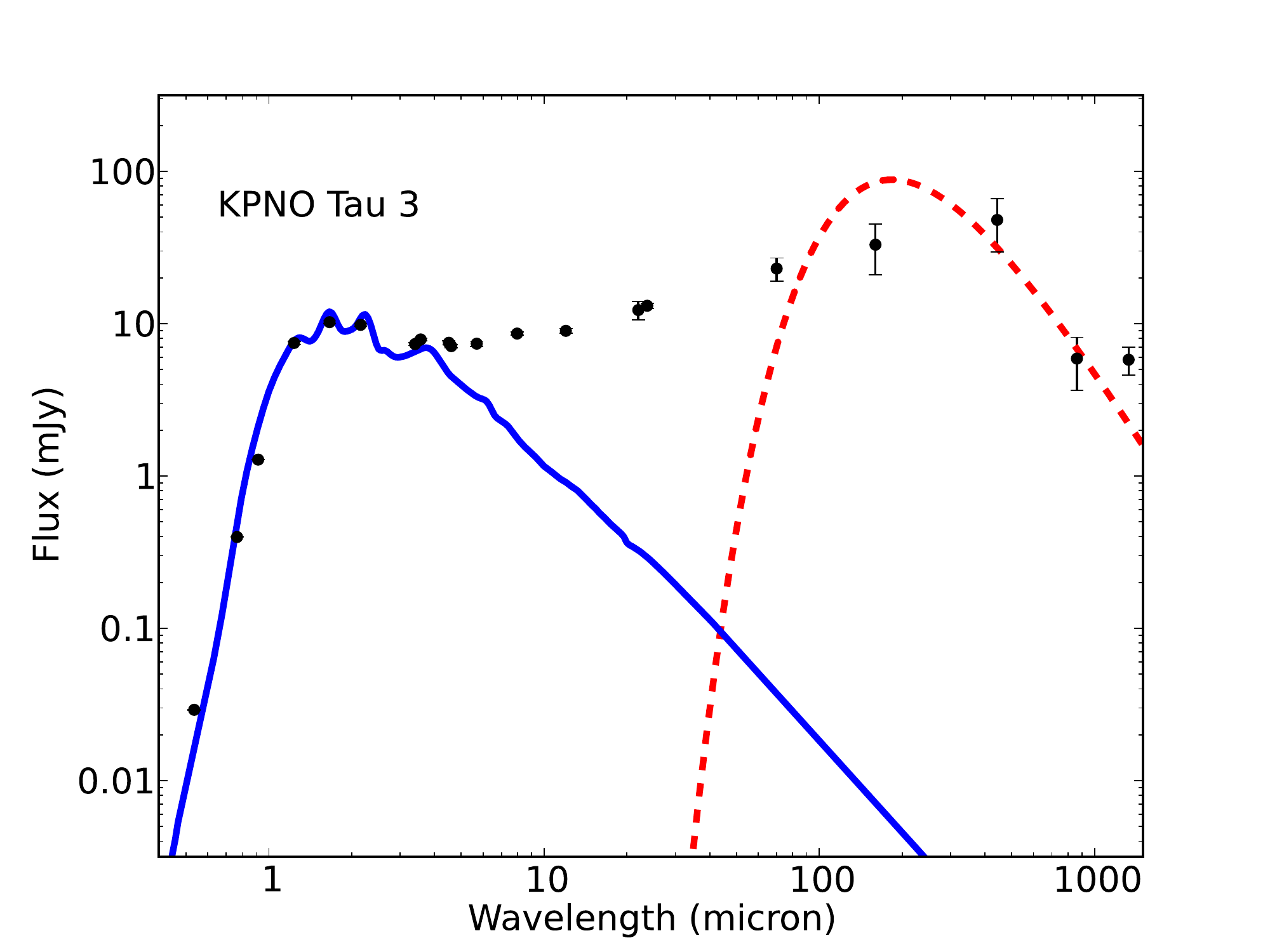}\\
\includegraphics[scale=0.45, clip=True, trim=0.4cm 0.2cm 2cm 1.2cm]{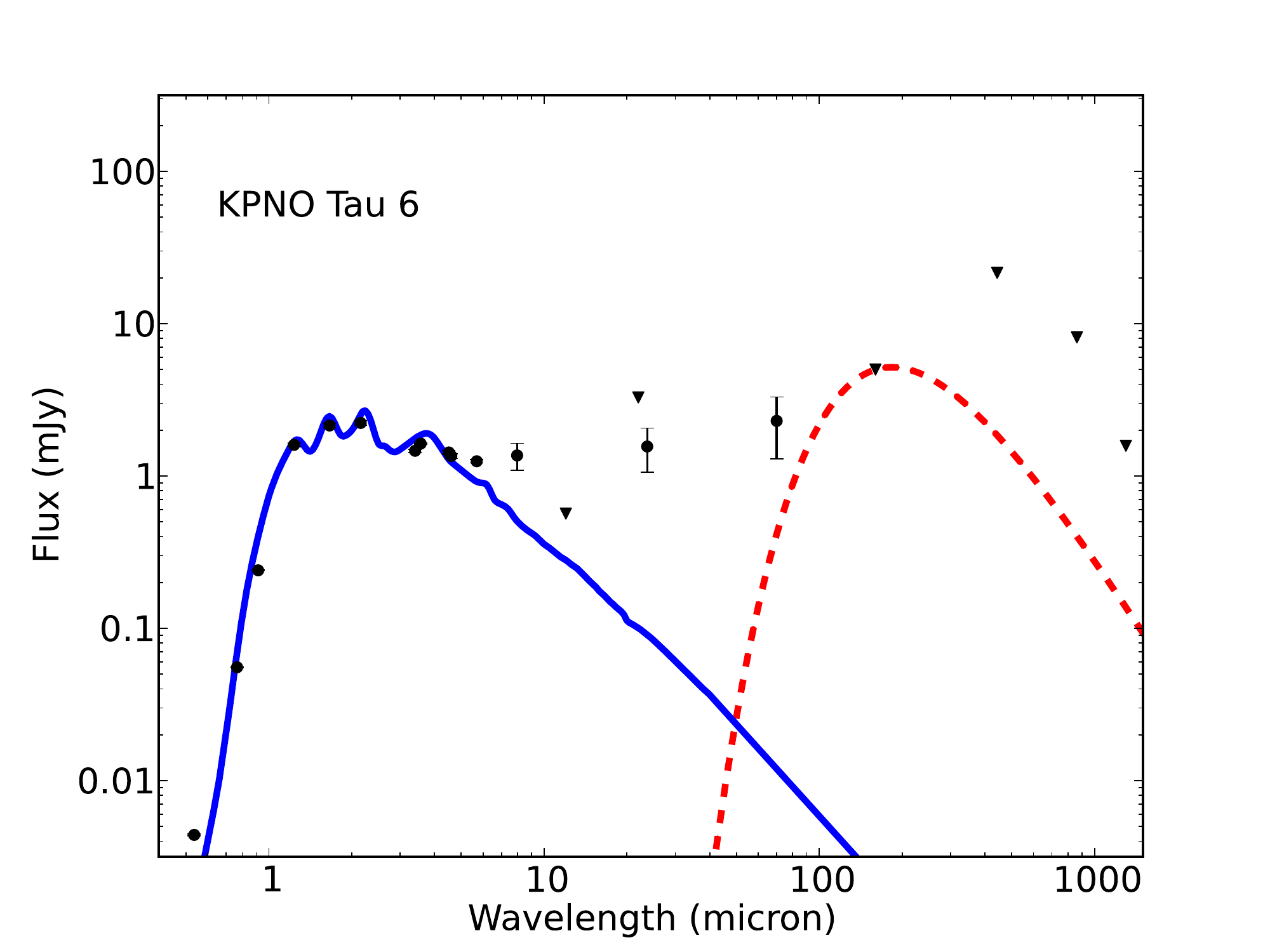}
\caption{\label{fig:fits}
SEDs for KPNO Tau 1, KPNO Tau 3, and KPNO Tau 6 (top to bottom). Circles mark observed photometry and triangles mark upper limits (\textit{HST}, 2MASS, \Spitzer, \textit{AKARI}, \Herschel, and SCUBA). The blue solid line shows the expected photospheric emission using NextGen models. The red dashed line traces the emission from the assumed 20 K dust used to measure disk mass. These lines correspond to $2.7\times10^{-4}$ \Msun, $4.0\times10^{-4}$ \Msun, and $2.3\times10^{-5}$ \Msun\ disks (according to Equation~(\ref{eq:dustMass})) for KPNO Tau 1, KPNO Tau 3, and KPNO Tau 6, respectively. Note that these disk masses for KPNO Tau 1 and KPNO Tau 6 correspond to upper limits. For KPNO Tau 6, we have plotted emission from $\sim$10 $\times$ less dust than suggested by the SCUBA flux upper limits to agree with the 160 \micron\ flux upper limit from \Herschel.\scriptsize}
\end{figure}

\begin{deluxetable}{cccc}
\tablewidth{0pc}
\tablecaption{Brown Dwarf Properties
\label{tbl:kt_dwarfs}}
\tablehead{
\ch{Target} 	& \ch{$M_*$\tnm{a}} 	& \ch{\Teff\tnm{b}}	& \ch{\Av\tnm{c}} \\ 
\ch{} 		& \ch{(M$_{\odot}$)}	& \ch{(K)} 			& \ch{(mag)}}
\startdata
KPNO Tau 1 	& 0.020 $\pm$ 0.010 	& 2600				& 3.5 \\ 
KPNO Tau 3 	& 0.077 $\pm$ 0.009 	& 3000 				& 3 \\
KPNO Tau 6 	& 0.021 $\pm$ 0.007 	& 2600				& ~~3		
\enddata
\tnt{a}{Brown dwarf masses are from \citet{KrausHillenbrand2009}.}
\tnt{b}{\Teff~are from \citet{Bricenoetal2002}}
\tnt{c}{\Av are determined from the photospheric modeling described in Section~\ref{sec:phot_kt3}.}
\end{deluxetable}

The photosphere of the brown dwarf is modeled to subtract the expected photospheric flux from the observed flux. Any excess emission is attributed to circumstellar dust. We use the grid of NextGen models \citep{Allard1997}, which include brown dwarfs with effective temperatures as low as $\sim$1500 K, to characterize the photospheric emission. (COND, AMES, DUSTY, and NextGen models all yielded similar results.) These models are produced with a variety of effective temperatures (\Teff), surface gravities (\logg), abundances, and alpha enhancements. We assume solar metallicity without alpha enhancement and a \logg\ of 3 (a typical value for brown dwarfs) for this analysis. We adopt effective temperatures, listed in Table~\ref{tbl:kt_dwarfs}, from studies of our targets based on spectroscopic data \citep{Bricenoetal2002} as this is a more accurate method to determine spectral types than the modeling of photometric data.

We use a $\chi^2$ minimization method on models with varying \Av\ to determine the optimal photospheric model by first normalizing the model to the $K_s$ flux. We then compare the \textit{Hubble Space Telescope (HST)} photometry \citep{Kraus2006} and 2MASS \citep{Cutri2003} photometry to the local average of the model at the effective observed wavelengths. The fluxes at these wavelengths are assumed to follow the photosphere (i.e., we do not expect any excesses in these bands). Table~\ref{tbl:kt_dwarfs} shows the parameters of the best fit photospheric  models. Figure~\ref{fig:fits} shows the SEDs for KPNO Tau 1, KPNO Tau 3 , and KPNO Tau 6, respectively.


\subsection{Determining Disk Masses}  
\label{sec:mass_kt3}

The 450 \micron\ and 850 \micron\ emission is assumed to originate in the optically thin cold dust that dominates the disk mass. The mass of the disk, $M_{disk}$, can be determined from the flux density of the dust at a given wavelength, $F_{dust}$, by
\begin{equation}
\label{eq:dustMass}
M_{disk} = \frac{F_{dust} D^2}{\kappa_{\nu}B_{\nu}(T)},
\end{equation}
where $D$ is the distance to the source, $\kappa_{\nu}$ is the opacity of the dust grains, and $B_{\nu}(T)$ is the Planck function for temperature, $T$. A temperature of 20 K is assumed and the opacity is assumed to be $\kappa_{\nu}=0.1(\nu/$1000~GHz$)$ cm$^2$ g$^{-1}$, following previous studies of the Taurus region \citep{AndrewsWilliams2005,Beckwithetal1990}. (This opacity relation includes an assumed dust-to-gas ratio of 1:100.) Although there may be systematic uncertainties in the assumptions on $\kappa_{\nu}$ and $T$, these assumptions are used for other studies of the Taurus star-forming region and therefore are valid for comparing our results to those found for classical T-Tauri stars in this region.

Disk masses for KPNO Tau 3 and upper limits for KPNO Tau 1 and KPNO Tau 6 are determined using the 450 \micron\ and 850 \micron\ fluxes and Equation~(\ref{eq:dustMass}). These measurements are listed in Table~\ref{tbl:kt_disks}. A disk mass of 4.0 $\times~10^{-4}$ \Msun\ for KPNO Tau 3 is determined using a $\chi^2$ fit to the SCUBA fluxes using the least-squares fitting package MPFIT \citep{Markwardt2009}. The relative disk-to-host mass for KPNO Tau 3 is then $\sim$0.5\% (and $<$1\% for KPNO Tau 1 and KPNO Tau 6). This relative disk mass is comparable to the values of $\lesssim$1\% and 5\% that have been found for other brown dwarfs and agrees with the values for low mass T Tauri stars \citep{ScholzJayaWood2006}.


\subsection{Column Density of \tco\ and \ceo\ toward KPNO Tau 3}
\label{sec:massMolecules_kt3}

The SED of KPNO Tau 3 suggests that it is a Class II object \citep{Luhmanetal2010}, a young substellar object whose circumsubstellar material is located in a disk. It is possible, however, that some Class II objects, classified by their SEDs, are actually Class I objects with a remnant envelope \citep{Evans2009}. In this scenario, the viewing angle causes the observed SED to resemble that of a Class II object. For this reason, we took observations of \ceo\ $J=2-1$, an effective tracer of dense material, to place constraints on the presence of a dense remnant envelope.

At $\sim$230 GHz, the JCMT has a beam efficiency of 0.60.
We use this factor to determine the brightness temperature, $T_B$, from the antenna temperature, $T_A^*$. The optical depth, $\Delta\tau_o$ is calculated using
\begin{equation}
\label{eq:tau}
T_B = T_o~[f(T_{\rm{ex}}) - f(T_{\rm{bg}})]~[1-\rm{exp}(-\Delta\tau_o)]
\end{equation}
where $T_o = h \nu_o / k$ and $f(T) = [\rm{exp}(T_o/T) - 1]^{-1}$, $\nu_o$ is the frequency at line center listed in Table~\ref{tbl:spectra}, $h$ is the Planck constant, and $k$ is the Boltzmann constant. The excitation temperature, $T_{\rm{ex}}$, is assumed to be equal to the kinetic temperature of the dust, 20 K, (discussed in Section~\ref{sec:mass_kt3}) assuming that the cloud is in local thermodynamic equilibrium. The background temperature, $T_{\rm{bg}}$, is that of the cosmic microwave background, 2.73 K. The emission lines from both isotopologues are found to be optically thin ($\Delta\tau_o << 1$) with $\Delta\tau_o$ of 0.050 and $<$0.0021 for \tco\ $J=2-1$ and \ceo\ $J=2-1$, respectively.

The column density, $N$, is given by
\begin{equation}
\label{eq:column}
N = {{8 \pi \nu_o^2 \Delta\nu Q \Delta\tau_o} \over {c^2 A_{21}}} ({ g_2 \over g_1}) [1 - e^{-T_o / T_{\rm{ex}}}]^{-1}
\end{equation}
where the partition function, $Q$, is  $\sim 2 T_{\rm{ex}}/T_o$. $g_2$ and $g_1$ are the statistical weights of the J=2 and J=1 rotational levels, respectively. $A_{21}$ is the Einstein coefficient for the 2--1 transition and has a value of $10^{-6.22}$ for both \tco\ and \ceo.\footnote{All spectral line data were taken from the Spectral Line Atlas of Interstellar Molecules (SLAIM) for C$^{18}$O (F. J. Lovas, private communication, \citealt{Remijanetal2007}) and the Cologne Database for Molecular Spectroscopy (CDMS) for $^{13}$CO \citep{Muelleretal2005} (available at http://www.splatalogue.net).}  $\Delta\nu$ is measured for the \tco\ line and assumed to be the same for the \ceo\ line.

The measured column density of \tco\ is $\sim$$(3.8 \pm 1.0) \times 10^{14}$ cm$^{-2}$ and that for \ceo\ is $\lesssim$$3 \times 10^{13}$ cm$^{-2}$ (2$\sigma$ upper limit). Were there an envelope around KPNO Tau 3, one would expect to find $N$(\tco)/$N$(\ceo) in the range of 10 -- 16, as is found for dark clouds \citep{Zhu2007,Kim2006}. A substantial amount of \ceo, such as would be present in the dense material of a circumsubstellar envelope, would yield a ratio lower than this range. We measure, however, a ratio of $N$(\tco)$/N$(\ceo) $\gtrsim$ 13, consistent with this range and therefore we do not suspect that a remnant envelope is present.


\section{SED modeling for KPNO Tau 3}
\label{sec:model}

\begin{figure*}
\includegraphics[width=6.0 in,clip=True, trim=0cm 0cm 0cm 0cm]{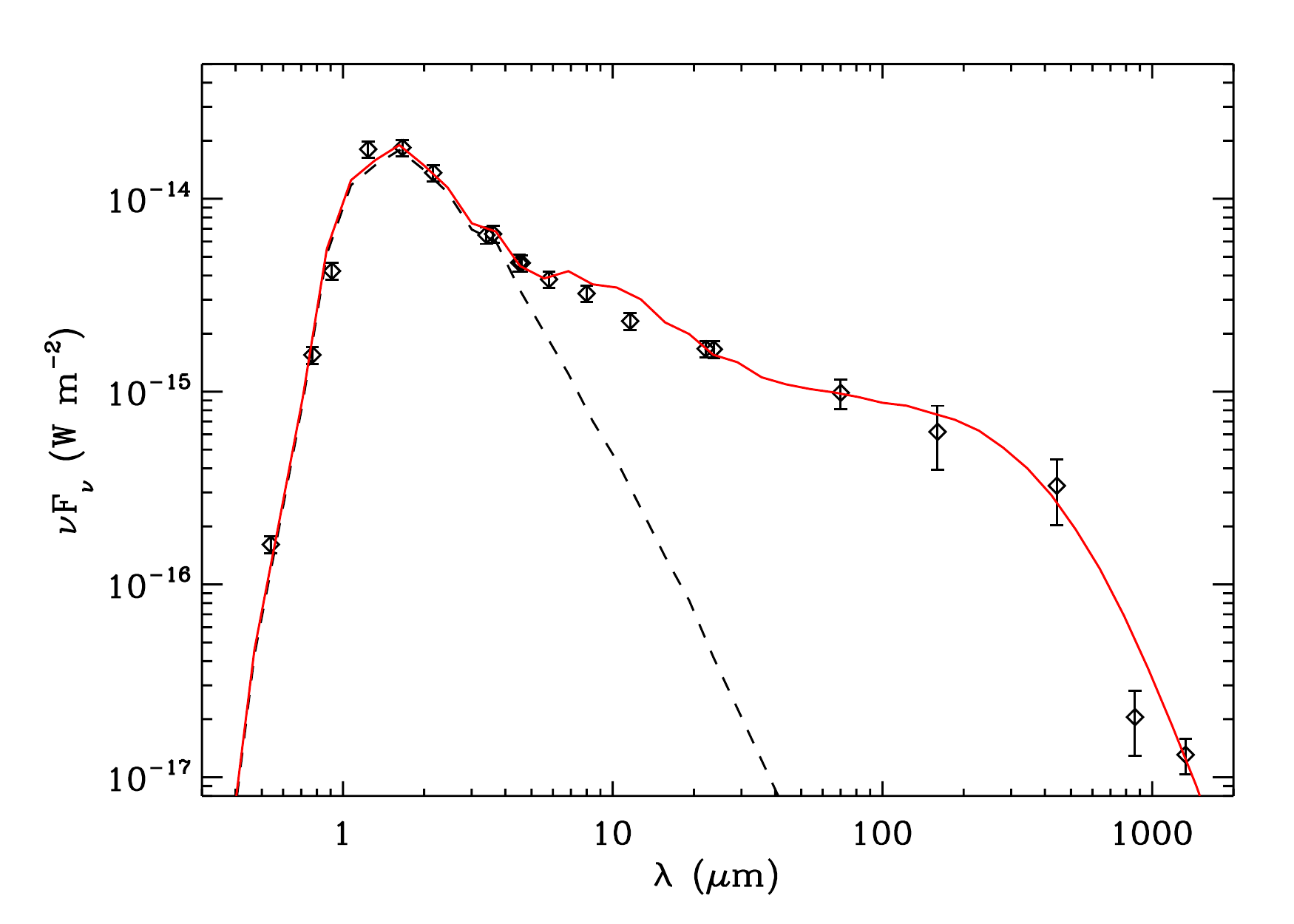}
\caption{\label{fig:model}Best fit photosphere and disk models for KPNO Tau 3 with a flared geometry. The black dashed curve traces the expected photospheric emission from the NextGen model. The red solid curve shows the model including the thermal dust emission from a broad distribution of grains from 0.03~\micron\ to $\sim$1 mm. These models have been fit to the observed SCUBA, \Herschel, and SMA photometry (black diamonds).\scriptsize}
\end{figure*}

\begin{deluxetable}{cccc}
\tablewidth{0pc}
\tablecaption{Fitted Parameters for KPNO Tau 3's disk\label{tbl:model}}
\tablehead{\ch{Parameter}	& \ch{Units}		& \ch{Fitted Range in Values\tnm{a}}	&\ch{Single Best Fit model\tnm{b}}}	
\startdata
\amin\tnm{c}  				& \micron		& 0.03								& 0.03 \\
\amax		  				& \micron		& $>$5.3 (1.1)						& 800 \\
$\eta$\tnm{c}  				& \nodata		& -3.5								& -3.5 \\
$\gamma$						& \nodata		& Unconstrained\tnm{e} 				& -1.4 \\
$R_{out}$		 			& AU				& $>$5.8 (2.9)						& 44 \\
$R_{in}$  					& AU				& $<$0.30 (0.65)						& 0.10 \\
$\beta$  					& \nodata		& $1.19^{+0.11}_{-0.08}$				& 1.21 \\
$H_0$  						& AU				& $9.9^{+5.6}_{-4.0}$				& 9.9 \\
$R_*$  						& R$_\odot$		& $0.56^{+0.09}_{-0.07}$				& 0.55 \\
$M_{dust}$\tnm{d} 			& M$_\odot$		& $>$4.4 (1.1) $\times 10^{-6}$		& \textcolor{white}{spc} 2.0 $\times 10^{-5}$
\enddata
\tnt{a}{Quoted limits are 2$\sigma$ (3$\sigma$).}
\tnt{b}{This is the model plotted in Figure~\ref{fig:model}. The single best fit model represents the model with the lowest \chisq\ value and its parameter values fall within the fitted parameter ranges in preceding column.}
\tnt{c}{The values of these parameters are fixed in the model.}
\tnt{d}{The total disk mass is assumed to be $\sim100 \times M_{dust}$.}
\tnt{e}{The power law index of the surface density law had a flat probability distribution across the fitted parameter range (-2 to 0).}
\end{deluxetable}

We detect submillimeter continuum emission toward KPNO Tau 3 and are able to derive a complete model for this target. We model the SED to constrain dust properties (dust mass, maximum grain size, surface density index), disk geometry (inner and outer radii, flaring index, scale height) and KPNO Tau 3's substellar radius, $R_*$. The model SED of the disk is computed using the radiative transfer code MCFOST \citep{Pinte2006,Pinte2009}. The model of the disk extends from $R_{in}$ to $R_{out}$ with a total dust mass of $M_{dust}$ and has a surface density varying with disk radius as a power law with index $\gamma$. The disk is modeled with a flared geometry described by the scale height of the disk at a distance $r$, $H(r)=H_0~(r/r_0)^\beta $, where $H_0$ is the scale height at $r_0$ = 100 AU. The dust is modeled using Draine's astronomical silicates with the size distribution of the grains given by $N(a) da =a^{\eta} da$ which represents the number of grains with sizes from $a$ to $a + da$. The value $\eta$ is fixed to -3.5 according to the Dohnanyi size distribution \citep{Dohnanyi1969} and the distribution spans from the minimum grain size, \amin, to the maximum grain size, \amax.

We use the affine-invariant ``ensemble'' MCMC method proposed by \cite{GoodmanWeare2010} to sample the parameter space more finely and smartly in situations where model parameters are correlated with one another, as is clearly the case for disk SED modeling. We determine a single best fit (where the \chisq\ is a minimum for all fitted parameters) as well as the range in parameter values (based on the Bayesian probability distribution functions for each parameter). We include the 70 \micron\ and 160 \micron\ fluxes of $23\pm4$ mJy and $33\pm12$ mJy, respectively, measured with the \Herschel\ \citep{Bulger2014subm} and the 1.33 mm flux of $5.8\pm1.2$ mJy from the SMA \citep{Andrewsetal2013}.

The range in fitted parameter values are listed in Table~\ref{tbl:model} along with the values for the single best fit model plotted in Figure~\ref{fig:model} against the observations. (The median is quoted for the 1$\sigma$ parameter range, and although the median and the value for the single best fit model agree within the uncertainty ranges, the two are often slightly offset from  each other.) Although the observed 850 \micron\ flux falls below the best fit model, it agrees within 3$\sigma$ of the predicted flux of the model. This modeling gives a more physical basis to assess the disk (dust) mass and allows for some geometrical parameters to be constrained.

The results of the SED fitting suggest the disk is moderately flared with a flaring index consistent with that measured for other brown dwarf disks and T-Tauri stars \citep{Harveyetal2012}. The surface density index is not strongly constrained and the scale height is nominal for a low-mass central object. The minimum disk outer radius is $\sim$6 AU (2$\sigma$). Note that this is a lower limit to outer radius and the dust could indeed extend out to the typical brown dwarf disk size of 100 AU. Our limit is also consistent, although less stringent, than the minimum outer radius of 20 AU required by \cite{Mohantyetal2013} to explain the observed submillimeter/millimeter fluxes of brown dwarf disks. 

The fitted lower limit on the maximum grain size suggests that at least \micron-sized grains are present in the disk. However, the best fit model and a large fraction of the models from the eMCMC process favor much larger grain sizes (\amax$>$100 \micron\ for $\sim$67\% of all models). Therefore our modeling favors substantial grain growth in this disk, although the relatively low signal-to-noise ratios of most datapoints beyond 50 \micron\ preclude a definite conclusion. It is difficult to expand the range of maximum grain sizes explored in the SED modeling as submillimeter/millimeter observations are not sensitive to cm-sized and larger grains, and therefore including them will skew the median of the probability distribution. 

The derived constraint on the disk mass, $>$$4.4 \times 10^{-4}$ \Msun, is consistent with the measured disk mass in Section~\ref{sec:mass_kt3}. This is expected since the opacity assumed in Section~\ref{sec:mass_kt3} is valid for dust distributions containing millimeter-sized grains. Similar to the argument for constraining the maximum grain size, the submillimeter/millimeter observations are not sensitive to cm and larger sized grains and therefore the disk mass contained in these larger dust grains.


\section{Conclusions}
\label{sec:conclusions}

Our submillimeter data show that KPNO Tau 3 has a significant cold component of dust suggesting a disk (gas + dust) mass of $\sim$$4.0 \times 10^{-4}$ \Msun. We have also placed upper limits on the disk masses for KPNO Tau 1 and KPNO Tau 6 of $<$2.1 $\times10^{-4}$ \Msun\ and $<$2.7 $\times10^{-4}$ \Msun, respectively. These upper limits are comparable to those reported by \cite{ScholzJayaWood2006} ($\sim$2.7 $\times 10^{-4}$ \Msun; using Equation~(\ref{eq:dustMass}) and our assumptions). In the case of KPNO Tau 6, this upper limit is consistent with the most probable disk mass of $10^{-5}$ \Msun\ determined by SED modeling and \Herschel\ observations by \cite{Harveyetal2012}, however a disk with our mass upper limit over predicts their measured 160 \micron\ flux upper limit. This means that we can exclude the presence of a large reservoir of large bodies in the disk midplane to which \Herschel\ would not be sensitive. However, as of yet there is no direct measurement of the dust mass in KPNO Tau 6's disk. 

We measure a submillimeter spectral index of $\alpha = 3.3 \pm 1.1$ for KPNO Tau 3 from its 450 \micron\ and 850 \micron\ fluxes, consistent with the value of $2.0 \pm 0.5$ measured for disks in Taurus by \cite{AndrewsWilliams2005}. We have confirmed that this is consistent with the spectral index between 450 \micron\ and 1.33 mm (using the flux from \citealt{Andrewsetal2013}) of $2.0 \pm 0.5$ for KPNO Tau 3, since the 850 \micron\ appears low in comparison to the 1.33 mm flux. This is unexpected given that interferometric observations tend to measure lower fluxes than single dish telescopes, as they filter out emission on larger spatial scales (such as emission from the cloud). Given the uncertainties in flux, it is not clear whether there is a break in the submillimeter slope, which would suggest a break in the grain size distribution. 

We establish that the dust detected towards \kt\ lies in a circumsubstellar disk. Some young objects with an envelope (Class Is) can have similar SEDs to young objects with only disks (Class IIs) if the opening angle of the envelope is directed towards us. This situation arises because young stellar object classification is typically based on the spectral slope in the near- and mid-infrared; without the long-wavelength information, it is difficult to extrapolate the SED or determine the wavelength range at which the majority of the energy is emitted. The submillimeter fluxes for \kt, however, reveal that its SED is characteristic of a Class II source with the majority of energy radiated at 1 -- 10 \micron.


The analysis of the disk around \kt\ is further enhanced by modeling the disk with a flared geometry. This model constrains the physical properties of the disk. In some cases we are only able to place upper or lower limits on the fitted parameters, as can be expected from the limitations of modeling the SED in the absence of resolved imaging. 
The modeling results suggest that the disk geometry is also similar to that for T Tauri stars and favors the presence of larger-sized (\amax\ $>$ 100 \micron) dust grains, but the data are not tight enough to completely exclude a relatively small \amax. The $\sim$3 Myr age of \kt\ \citep{Barrado2004} supports the indication of grain size evolution, given that evidence of dust grain growth has been observed in other brown dwarf disks, and that large grains can grow in such disks if they are scaled down versions of T Tauri disks as they have similar collisional timescales \citep{MeruGalvagniOlczak2013}. The SED modeling also confirms the simple mass estimates based solely on the submillimeter fluxes determined in Section~\ref{sec:mass_kt3}.

The presence of cold, 20 K dust implies that significant amounts of dust are present at large radii. This population of cold dust favors a formation mechanism for brown dwarfs that is similar to that of stars, since the ejection of a stellar embryo could truncate the disk. Furthermore, the amount of cold dust in the \kt\ disk is suggestive of the system being a lower-mass analogue of a T-Tauri star. This congruence is shown by considering the relative disk mass for \kt\ of 0.5\% (limits of $<$1\% for KPNO Tau 6 and KPNO Tau 1). This value is consistent with disk-to-host mass ratios for brown dwarfs and comparable to the range found for low-mass T-Tauri stars, $\lesssim$1\% -- 5\% (\cite{Mohantyetal2013}: see their Figure 3). The location of the cold dust can be investigated well with the Atacama Large Millimeter/Submillimeter Array (ALMA), since any remaining emission on larger spatial scales could be filtered out. ALMA would also be capable of measuring or placing constraints on disk sizes. Such data would constrain where the cold dust lies and determine whether or not the disks are likely to be truncated.

\acknowledgments

H.B.F. and B.C.M. acknowledge a Discovery Grant from the Natural Science \& Engineering Research Council (NSERC) of Canada.
We thank Gary Davis for awarding Director's Discretionary Time for the RxA3 data.


\bibliographystyle{apj}
\bibliography{library} 

\end{document}